\documentclass[prl,twocolumn]{revtex4-1}

\usepackage{enumerate}
\usepackage{amsmath,amsthm,amssymb}
\usepackage{graphicx}
\usepackage{listings}
\usepackage{color}
\usepackage{braket}

\newcommand{\mto}[0]{MgTi$_2$O$_4$} 

\newcommand{\up}[0]{\uparrow}
\newcommand{\dn}[0]{\downarrow}

\begin{document}
\title{Prediction for the singlet-triplet excitation energy for the spinel MgTi$_2$O$_4$ using first-principles diffusion Monte Carlo}
\author{Brian Busemeyer, Gregory J. MacDougall, Lucas K. Wagner}
\affiliation{Department of Physics, University of Illinois at Urbana-Champaign, Urbana, IL 61801} 
\begin{abstract}
  The spinel, MgTi$_2$O$_4$, undergoes a transition into a dimerized state at low temperatures that is expected to be a spin singlet.  
  However, no signature of a singlet-triplet transition has been discovered, in part due to the difficulty of predicting the energy of the transition from theory. 
  We find that the dimers of {\mto} can be described by a Heisenberg model with very small interactions between different dimers. 
  Using high-accuracy first-principles quantum Monte Carlo combined with a novel model-fitting approach, we predict the singlet-triplet gap of these dimers to be 350(50)~meV, a higher energy than previous experimental observations have considered.  
  The prediction is published in advance of experimental confirmation.
\end{abstract}
\maketitle


\section{Introduction}

  Exotic electronic states are predicted to occur at low temperatures in geometrically frustrated lattices, including the pyrochlore lattice\cite{radaelli_orbital_2005,lee_frustrated_2010}. 
  However, as temperature is decreased, the frustration is often resolved by structural distortions.
  Because these systems have strong electron-electron interaction effects, the resultant ground states can be challenging to describe using simple electronic structure theories. 
The causation of this distortion is not always completely understood, and the correlated nature of the material makes theoretical approaches challenging.

  In the case of pyrochlore {\mto}, the high temperature pyrochlore lattice distorts to a dimerized structure at $T_s=260$~K\cite{schmidt_spin_2004}. 
  A spin-singlet state would not be surprising in this case, since there is a sharp decrease in the magnetic susceptibility coincident with the dimerization\cite{isobe_observation_2002}.
  Existence of the singlet ground state is also indirectly supported by optical experiments\cite{zhou_optical_2006} interpreted through density functional theory (DFT) calculations with the B3LYP functional\cite{schmidt_spin_2004}.
  Local spin density approximation (LSDA) + $U$ calculations find that at nonzero $U$, the ground state contains antiferromagnetic dimers\cite{leoni_orbital-spin_2008}.
  However, both generalized gradient approximation (GGA) and local density approximation (LDA) + $U$ with $U < 6$~eV fail to produce the insulating state, highlighting the dependence of the results on the choice of DFT functional.
  An additional concern with the spin-singlet identification is that neutron scattering experiments of {\mto} find no singlet-triplet excitation up to energies of 25~meV\cite{lee_frustrated_2010}.
  We are not aware of any attempt to predict the singlet-triplet gap of the system from first-principles calculations.

  Another system, VO$_2$, is quite similar to {\mto} in that it undergoes a metal-insulator transition at low temperatures simultaneously with a structural transition that forms dimers.
  An inelastic x-ray scattering measurement found the singlet-triplet gap of VO$_2$ to be 460~meV\cite{he_measurement_2016}.
  Diffusion Monte Carlo (DMC) was applied to compute the singlet-triplet gap of VO$_2$, finding the gap to be 440(24)~eV---within statistical errors of the experiment\cite{zheng_computation_2015,zheng_erratum:_2018}.
  The analysis used the DMC calculations to fit a model Heisenberg Hamiltonian, which can be solved for the excitation energies.
  Similar techniques have been successfully applied widely in the literature, including in calculations utilizing quantum Monte Carlo\cite{reinhardt_magnetic_1999,de_graaf_electronic_2000,duan_magnetic_2006,ma_arsenic-bridged_2008,mazin_problems_2008,akamatsu_antiferromagnetic_2011,seo_strain-driven_2012,foyevtsova_ab_2014,zheng_computation_2015}.
  Recently, the method was generalized and given the theoretical grounding to fit a large family of models, including Hubbard and generalized Hubbard models\cite{zheng_real_2018}, which would give access to additional types of excitation energies.
  However, to cleanly demonstrate its predictability, correct predictions with the method should be published before the experimental observation.

  In this manuscript, we use quantum Monte Carlo calculations and a rigorous theory of effective models\cite{zheng_real_2018} to make a prediction for the singlet-triplet gap of {\mto}. 
  The gap is predicted to be 350(50)~meV, which should be observable using neutron scattering or reasonant inelastic x-ray scattering (RIXS) experiments. 
  Observation of an excitation at the singlet-triplet gap would be strong evidence for the singlet ground state, and confirmation that our technique is a new predictive tool for calculation of excited state properties in transition metal systems.

\section{Method}

  Our approach utilizes diffusion Monte Carlo (DMC) to provide energies as inputs into a model fitting approach, both of which are summarized here.

  Diffusion Monte Carlo \cite{foulkes_quantum_2001} is a stochastic projection method for finding the ground state of a Hamiltonian, $H$.
  It projects the ground state, $\ket{\Psi_0}$,  of $H$ from a trial wave function, $\ket{\Psi_T}$, by noting that
  $$
    \lim_{\tau \to \infty}
    e^{-\tau H} \ket{\Psi_T} 
    =
    \ket{\Psi_0}.
  $$
  In our work, $H$ is the first-principles Hamiltonian,
  $$
    H 
    =
    \frac 12 \sum_i \nabla_i^2
    +
    \sum_{ij} \frac{1}{r_{ij}}
    +
    \sum_{iI} V_\text{eff}(r_i-R_I),
  $$
  where $i$ indexes the electrons, $r_i$ are electron positions, $R_I$ are the ion positions, and $V_\text{eff}$ is a small-core pseudopotential provided by Burkatzki, Filippi, and Dolg (BFD)\cite{burkatzki_energy-consistent_2007,burkatzki_energy-consistent_2008}.
  By stochastically applying the projection operator, $e^{-\tau H}$ to $\ket{\Psi_T}$, it exponentially suppresses all but the ground state component of $\ket{\Psi_T}$.
  Without further approximation, the method suffers from a sign problem when applied to antisymmetric wave functions, such as in many-fermion systems.

  The fixed-node approximation avoids this sign problem by constraining the projected wave function, $\lim_{\tau \to \infty} e^{-\tau H} \ket{\Psi_T}$, to have the same nodes as $\ket{\Psi_T}$.
  This fixed-node wave function, $\ket{\Psi_\mathrm{FN}}$ is the lowest energy wave function with a given set of nodes.
  If the trial function nodes are exactly the ground state nodes, then the method remains an exact ground state method.
  If the trial function nodes are approximately the ground state nodes, it will suffer some error known as fixed-node error, which depends only on the accuracy of the nodes of $\ket{\Psi_T}$.
  Similarly, if the trial function nodes are approximately the excited state nodes, in practice, the method can give reasonably accurate estimates for excited state energies\cite{hunt_quantum_2018}.
  The accuracy of fixed-node DMC has been established in a range of transition metal and other strongly correlated systems\cite{wagner_energetics_2007,foyevtsova_ab_2014,yu_towards_2015,santana_structural_2015,mitra_many-body_2015,wagner_discovering_2016,doblhoff-dier_diffusion_2016,santana_diffusion_2017}.

  We used a conventional recipe to perform the DMC calculations.
  To generate the trial functions, we employed single-Slater-Jastrow-type wave functions\cite{wagner_qwalk:_2009,needs_continuum_2010,kolorenc_wave_2010,shulenburger_quantum_2013}. 
  The Slater determinant is provided by DFT calculations, which were performed by the CRYSTAL package\cite{dovesi_quantum-mechanical_2018}.
  In previous studies of hybrid functionals and fixed-node error, the PBE0 functional often provides the best nodes compared to other choices of common DFT functionals\cite{kolorenc_wave_2010,wagner_effect_2014,zheng_computation_2015,busemeyer_competing_2016}, so we employed the PBE0 exchange-correlation functional for the DFT calculations in this work.
  We used a local Gaussian triple-$\zeta$ with polarization (TZP) basis set for the PBE0 calculations.
  The Jastrow factor was energy-optimized with variational Monte Carlo within the QWalk package\cite{wagner_qwalk:_2009,umrigar_alleviation_2007}.
  Fixed-node diffusion Monte Carlo calculations were performed using QWalk, using $T$-moves to handle the nonlocal parts of the pseudopotential\cite{casula_beyond_2006}.
  The timestep error was confirmed to be negligible for these quantities (Supplemental Information).
  All DMC energies were twist-averaged over the 8 real twist-boundary conditions.

  We fit a nearest-neighbor Heisenberg model to our DMC data, and use it to calculate the singlet-triplet gap.
  \begin{align}
    \label{eq:heisenberg}
    H_\mathrm{eff}
    &=
    J \sum_{\langle i j \rangle} S_i \cdot S_j
  \end{align}
  where $\langle i j \rangle$ represent the pairs of Ti in a dimer.
  Because each dimer is independent, the Hamiltonian is block diagonal with blocks corresponding to each dimer.
  Let $\ket{\up \up}$ represent the spin state on a single dimer with both spins up, and $\ket{\up \dn}$ be the corresponding state with one spin up and one spin down, etc.
  Although not all of them are eigenstates, the expectation value of the energies of these states according to (\ref{eq:heisenberg}) are $-J/4$ for $\ket{\up\dn}$ and $\ket{\dn\up}$ and is $J/4$ for $\ket{\up\up}$ and $\ket{\dn\dn}$.
  Thus, knowing the correct difference between the expectation value of the energies of $\ket{\up \up}$ state and the $\ket{\up \dn}$ state determines $J$.
  Similar approaches have been utilized in many systems to determine effective exchange couplings\cite{reinhardt_magnetic_1999,de_graaf_electronic_2000,duan_magnetic_2006,ma_arsenic-bridged_2008,mazin_problems_2008,akamatsu_antiferromagnetic_2011,seo_strain-driven_2012,foyevtsova_ab_2014,zheng_computation_2015}.
  The singlet-triplet gap of this Hamiltonian is then
  \begin{align*}
    \bra{\up\up} H_\mathrm{eff} \ket{\up \up} - \bra{s} H_\mathrm{eff} \ket{s}  = J/4 - (-3J/4) = J,
  \end{align*}
  where $\ket{s} = (\ket{\up \dn} - \ket{\dn \up})/\sqrt{2}$ is the singlet ground state.

\begin{figure}
    \begin{tabular}{ccc}
      triplet
      &
      dimer
      &
      inter-dimer 1
      \\
      \includegraphics[width=0.14\textwidth]{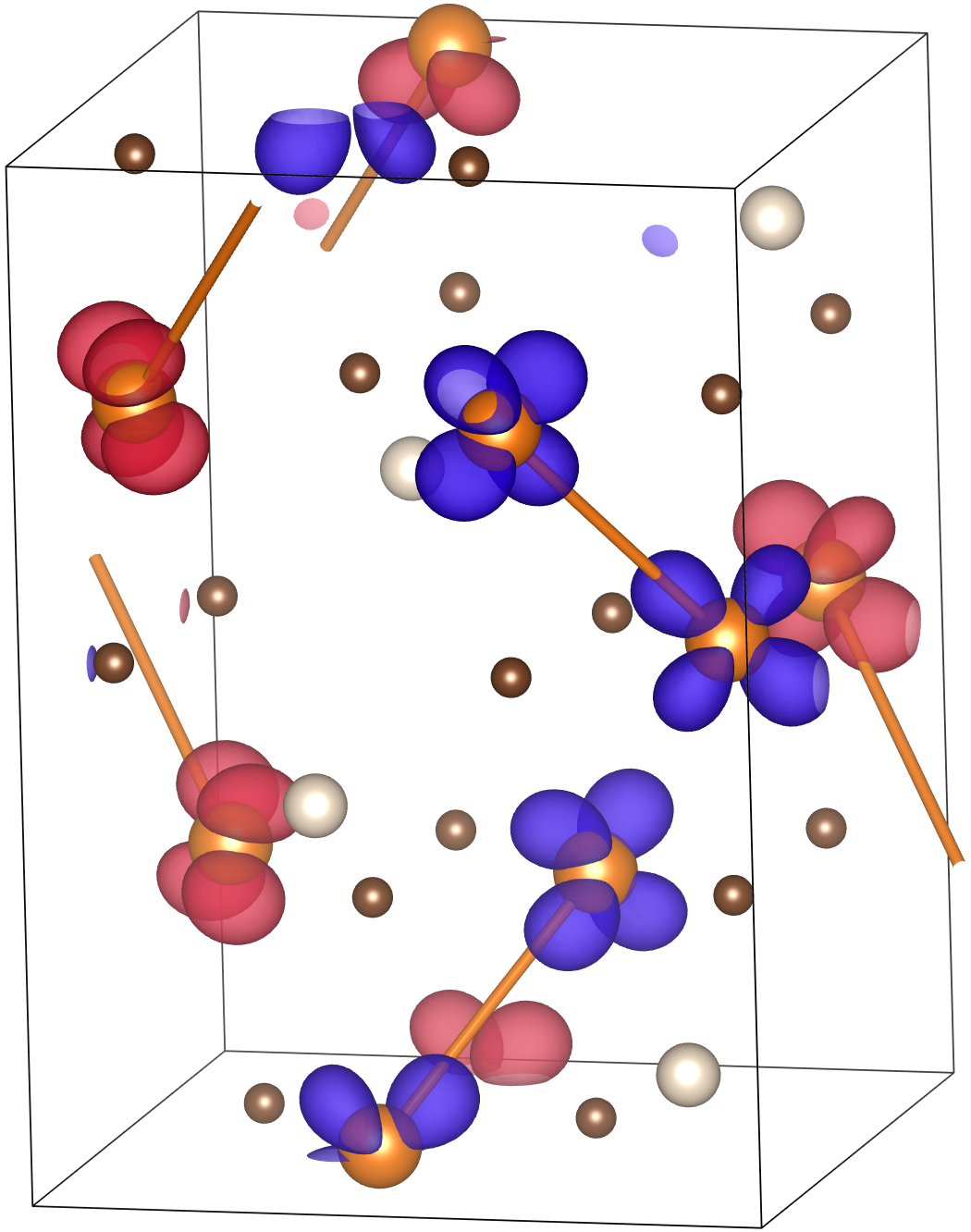}
      &
      \includegraphics[width=0.14\textwidth]{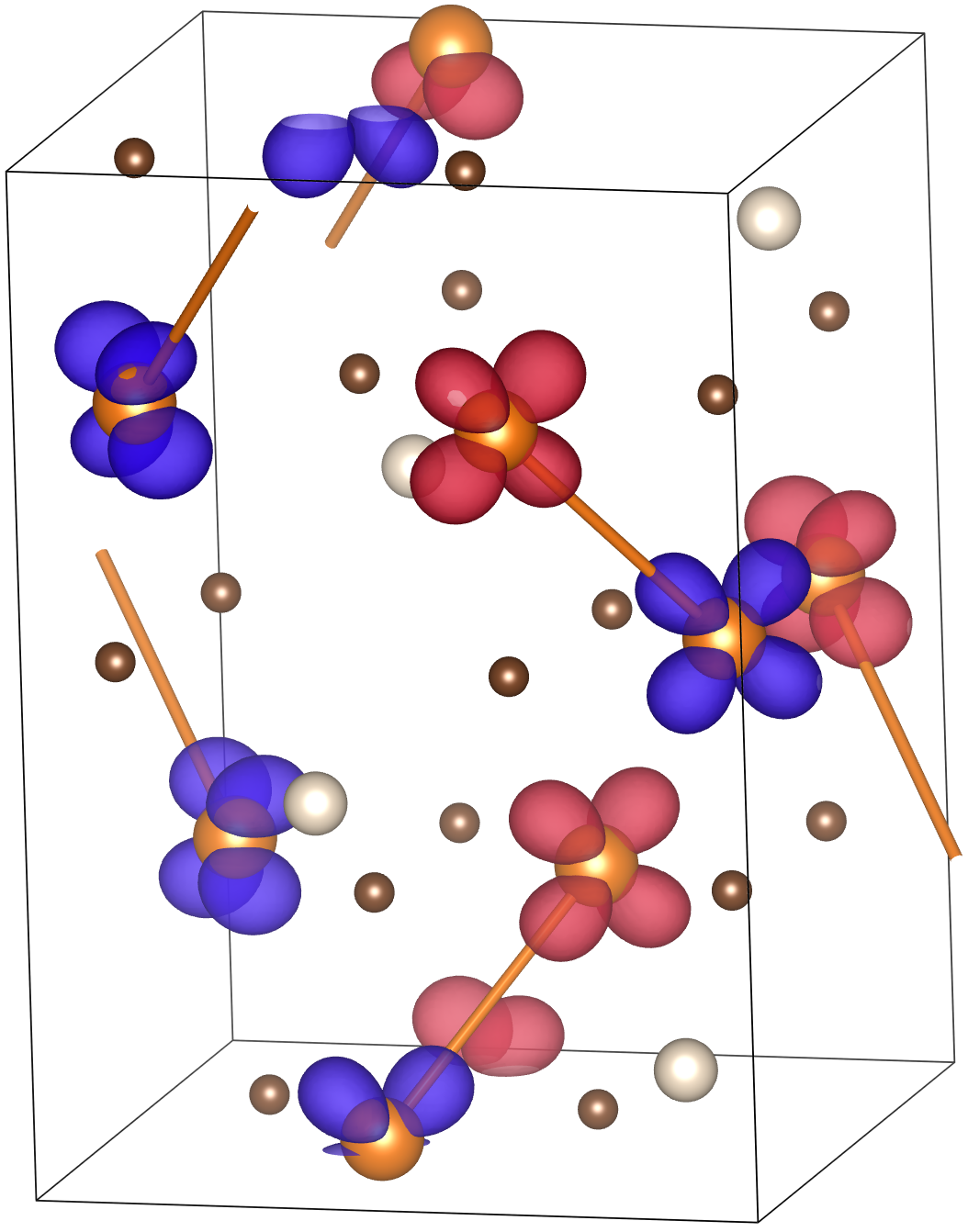}
      &
      \includegraphics[width=0.14\textwidth]{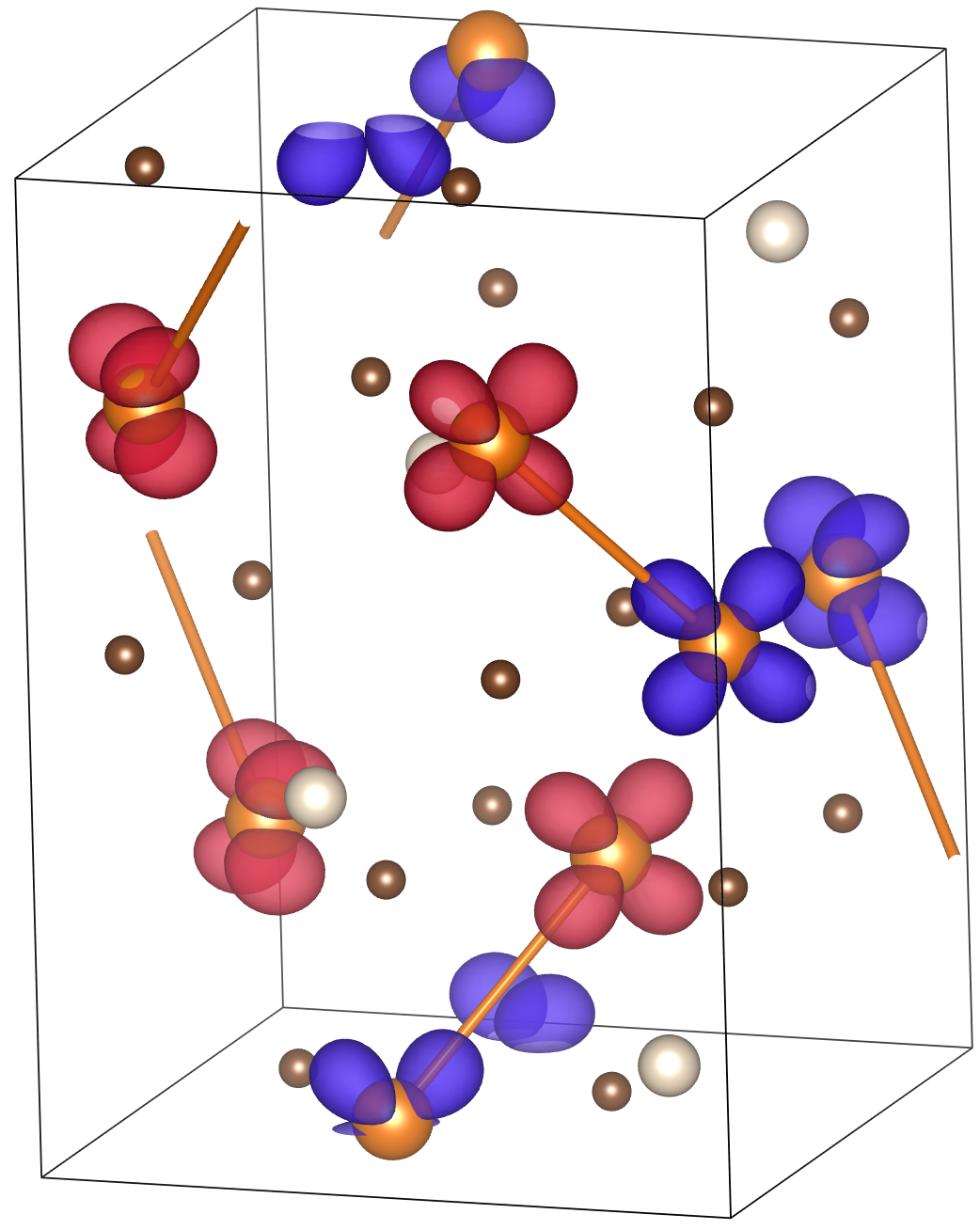}
    \end{tabular}
    \caption{
      \label{fig:contour}
      Spin and charge density contour plot for three spin states considered in this work.
      Up and down spin densities are represented as blue and red.
      Orange atoms with spin are Ti atoms, of the spinless atoms, the smaller brown atoms are O, and the white atoms are Mg. 
      Bonds are drawn between dimerized Ti atoms, and bonds that exit the unit cell have no Ti atom on the end outside the cell.
      The diffusion Monte Carlo energies of these states determine a Heisenberg model which we solved to compute the singlet-triplet gap.
    }
  \end{figure}

  Fixed-node DMC provided the estimated energy difference between $\ket{\up \up}$ and $\ket{\up \dn}$ state for each dimer.
  Beginning with a trial wave function with spin densities shown in Fig.~\ref{fig:contour}, the fixed-node constraint fixes the overall spin symmetry as the DMC projection removes all the high energy components from the trial wave function.
  Thus the resulting fixed-node wave function represents low-energy states with the spins on each Ti corresponding to $\ket{\up \up}$ or $\ket{\up \dn}$ on each dimer.


\section{Results}

\begin{figure}
    \begin{center}
      \includegraphics[width=0.5\textwidth]{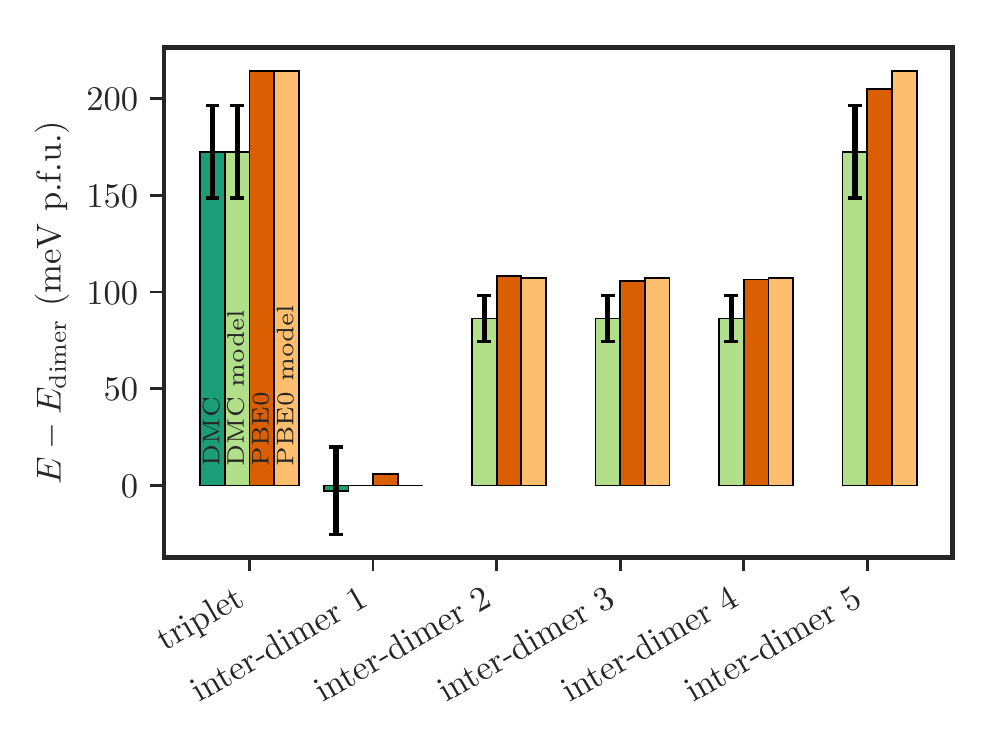}
      \caption{
        \label{fig:energy}
        Energies of each state relative to the dimer state of Fig.~\ref{fig:contour} according to DMC and PBE0 as well as models fit to DMC and PBE0 data.
        The triplet energy difference is sufficient to fit the parameter in (\ref{eq:heisenberg}), with $J = 350(50)$~meV from DMC, for example.
        This makes the singlet-triplet gap 350(50)~meV for this system.
        The inter-dimer states are predictions from the independent dimer model~(\ref{eq:heisenberg}), and are used to test the predictability of the model.
        The model predictions for the inter-dimer states agree quite well with~(\ref{eq:heisenberg}), suggesting that the independent-dimer model is an accurate effective model for the spins in this system
      }
    \end{center}
  \end{figure}

  We evaluated the energetics of the ``dimer'' and ``triplet'' spin configurations presented in Fig.~\ref{fig:contour} to fit the dimer model~(\ref{eq:heisenberg}).
  We also evaluated the energy of the spin-unpolarized state, finding it to be 810(30)~meV higher than the ``dimer'' according to DMC and 898~meV according to PBE0.
 Net spin moments around each Ti are therefore energetically favorable in our calculations.
  The relative energy of the ``triplet'' state to the ``dimer'' state according to both PBE0 and DMC are presented in the ``triplet'' group of in Fig.~\ref{fig:energy}.
  Requiring the model~(\ref{eq:heisenberg}) to reproduce this relative energy constrains its parameter to be $J = 350(50)$~meV.
  Thus the singlet-triplet gap is predicted to be 350(50)~meV.

  To test the accuracy of the independent-dimer Heisenberg model~(\ref{eq:heisenberg}), we compare its predictions for an additional test-set of states.
  An important approximation of the independent-dimer model is that it neglects interactions between different dimers.
  The Ti-Ti distance within a dimer is approximately 2.85~{\AA}.
  There are two bonds that are 3.00~{\AA}, two bonds that are 3.01~{\AA}, and one long bond that is 3.16~{\AA}.
  Considering the bond lengths, the intra-dimer interaction should be the strongest, with possible corrections due to inter-dimer interactions.
  The importance of these intermediate-distance inter-dimer interactions was checked by sampling additional ``inter-dimer'' states with different relative dimer orientations as well as different numbers of parallel and antiparallel intra-dimer alignment. 
  For example, the state labeled ``inter-dimer 1'' in Fig.~\ref{fig:contour} will probe inter-dimer interactions, as it changes the relative orientations of the dimers.
  For the left three states in Fig~\ref{fig:energy} and the unpolarized state, PBE0 tends to reproduce the DMC energy differences within 2-3$\sigma$ ($\sigma$ being the statistical uncertainty of the DMC results), so we relied on the PBE0 estimates for most of the inter-dimer states.
  The independent-dimer model predictions appears in Fig.~\ref{fig:energy} along with the first-principle energy differences according to PBE0.
  The independent-dimer model can be fit to either the PBE0 or DMC triplet energy difference, and both resulting model predictions are presented in Fig.~\ref{fig:energy}.
  The $J$ from the independent-dimer model can be fitted to either the PBE0 or DMC energy difference, but their results agree within 2$\sigma$.
  The model predictions and PBE0 reference energies agree within 10 meV, so the energy of these spin states depends on intra-dimer interactions to high accuracy.
  Thus, the independent dimer model~(\ref{eq:heisenberg}) seems to be an accurate minimal model for the spin degree of freedom of {\mto}.

\begin{figure}
    \includegraphics[width=0.5\textwidth]{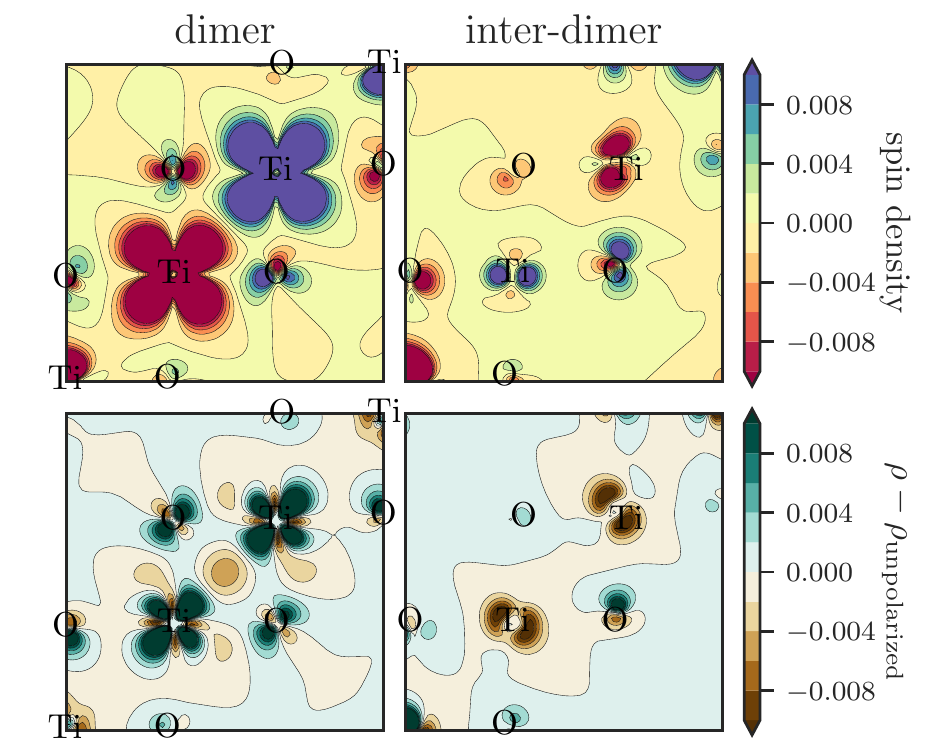}
    \caption{
      \label{fig:slices}
      Spin density (top row) and charge density ($\rho$) relative to the unpolarized charge density $\rho_\mathrm{unpolarized}$ (bottom row) through Ti pairs and their intermediate O.
      Dimer signifies the Ti atoms in the center form a nearest neighbor dimer pair in the structure.
      Inter-dimer signifies that the Ti atoms are nearest neighbors in the $x$-$y$ plane. 
      The dimer bond shows spin density typical for atoms interacting through superexchange, as well as an increase in hybridization between Ti and O within the dimer as compared to the unpolarized state.
      The inter-dimer bond shows no such signals of superexchange, in agreement with the model results.
    }
  \end{figure}

  The spin and charge density of the lowest energy state are shown in Fig~\ref{fig:slices}.
  The spin and charge density in DMC is within statistical uncertainties of the Slater determinant of PBE0 orbitals, hence we show the trial function densities to avoid the statistical noise in DMC.
  The 2-d slices, shown in Fig.~\ref{fig:slices}, are through nearest neighbor Ti atoms (intradimer) and through nearest neighbor Ti atoms in the $x$-$y$ plane (interdimer).
  The spin density in the dimer bond is typical for systems interacting through superexchange; in particular, the spin density on the oxygen tends to be opposite the spin of the nearest Ti atom.
  In contrast, the interdimer pairs ligands' spin densities do not show any obvious superexchange pattern.
  There is an increase in Ti-O-Ti hybridization relative to the spin-unpolarized state, but only inside the dimer bond, as one can see from the increased charge density.
  The spin and charge densities show little increase in hybridization outside the dimers, consistent with the results from fitting the models above.

\section{Conclusion}

  We have found the singlet-triplet gap of {\mto} using high-accuracy diffusion Monte Carlo calculations to be 350(50)~meV.
  The approach utilizes a downfolding approach that had been applied successfully to VO$_2$ and a variety of other systems\cite{zheng_real_2018}.
  A second benefit of this approach is that it can evaluate the accuracy of minimal models, which can provide simplified views of complex materials.
  In this case, we found that a Heisenberg model with interactions between Ti inside a dimer but with independent dimers can accurately describe the low-energy spin states in the system.
  Experimental confirmation of the singlet-triplet gap should be possible in RIXS or neutron scattering.
  We are publishing this result in advance of an experimental measurement so that it constitutes a true prediction. 
  If confirmed, this would be an example of a rare event in correlated electron physics--the prediction of a precise experimental result with quantitative accuracy, and cement diffusion Monte Carlo as a technique that can accurately predict properties of correlated electron materials.

\section{Acknowledgements}

  This research is part of the Blue Waters sustained-petascale computing project, which is supported by the National Science Foundation (awards OCI-0725070 and ACI-1238993) and the state of Illinois. Blue Waters is a joint effort of the University of Illinois at Urbana-Champaign and its National Center for Supercomputing Applications.

\section{Supplemental information}

\begin{figure}
  \includegraphics[width=0.5\textwidth]{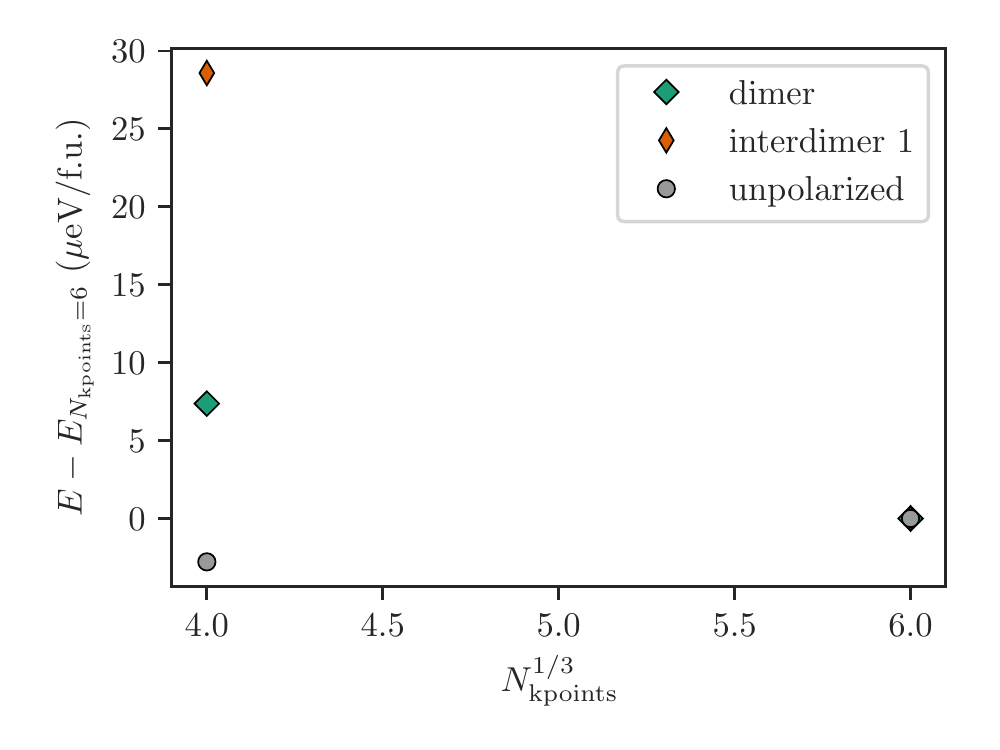}
  \caption{\label{fig:kpoint}
    Energy vs. Brillouin zone sampling for the PBE0 calculations.
    The energies are shown relative to the largest number of kpoints ($6 \times 6 \times 6$), so that the left point is zero by definition. 
    The error in the total energies is converged to the micro-eV scale, which is well below the scale of other errors in our work.
  }
\end{figure}

  We converged the Brillioun zone sampling in the PBE0 calculations in Fig.~\ref{fig:kpoint}.
  The total energies are already converged within the micro-eV scale with a  $4 \times 4 \times 4$ grid, which is well below the scale of the energy differences (hundreds of meV).

\begin{figure}
  \includegraphics[width=0.5\textwidth]{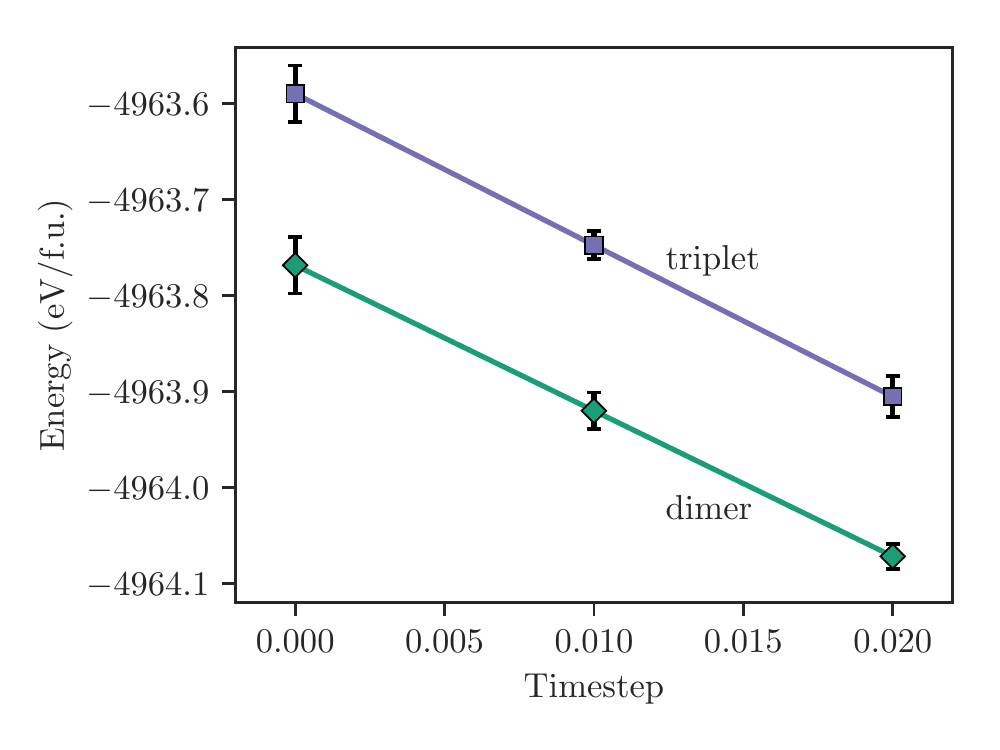}
  \caption{\label{fig:timestep}
    Timestep extrapolation for the DMC calculations was performed with two points at 0.01 and 0.02 timestep. 
    These points were extrapolated to zero timestep.
    The timestep error in the energy difference is well within our statistical error.
  }
\end{figure}

  We converged the timestep error from the DMC simulations in Fig.~\ref{fig:timestep}. 
  While the timestep error on the total energy is larger than the statistical error, this error completely cancels in any energy differences.
  The difference in extrapolated DMC energies match the difference in the 0.01 timestep energies within one standard error.

\bibliographystyle{unsrt}
\bibliography{mgti2o4.bib}
\end{document}